\documentclass[twocolumn]{aastex701}
\usepackage{graphicx}
\usepackage{CJKutf8}
\usepackage{url}
\graphicspath{{figure/}}
\usepackage{subcaption}

\begin{document}

\title{Multi-band power color-color diagrams of three black hole X-ray binaries observed with {\it Insight}-HXMT}

\author[0009-0003-4182-687X]{Shao-Feng Liu}
\affiliation{School of Physics and Astronomy, Sun Yat-sen University, Zhuhai 519082, People’s Republic of China}
\email{liushf37@mail2.sysu.edu.cn}

\author[0000-0001-9599-7285]{Long Ji}
\affiliation{School of Physics and Astronomy, Sun Yat-sen University, Zhuhai 519082, People’s Republic of China}
\affiliation{CSST Science Center for the Guangdong-Hong Kong-Macau Greater Bay Area, Zhuhai 519082, People’s Republic of China}
\email{jilong@mail.sysu.edu.cn}
\correspondingauthor{Long Ji}

\author[0000-0003-3207-5237]{Yanan Wang}
\affiliation{Key Laboratory of Optical Astronomy, National Astronomical Observatories, Chinese Academy of Sciences, 20A Datun Road, Beijing 100101, China}
\email{wangyn@bao.ac.cn}

\begin{abstract}
% {\color{blue} 
% {\bf 
% \begin{CJK*}{UTF8}{gbsn}(摘要最后改)
% \end{CJK*}
% }}
Power color-color diagrams (PCCDs) provide a useful diagnostic tool for studying the evolution of outbursts in black hole X-ray binaries.
In this paper, we present power color-color diagrams of three sources (MAXI J1348--630, MAXI J1820+070 and Swift J1727.8--1613) observed with {\it Insight}-HXMT in a wide energy range of 2--80 \,keV.
We compared the hue regions defined by {\it RXTE}, which are associated with different spectral states, with the {\it Insight}-HXMT results.
We find that, for hard and hard-intermediate states, the trajectories of {\it Insight}-HXMT in the power color-color diagrams are generally consistent with those of {\it RXTE}. 
In the soft and soft-intermediate states, weak variability generally prevents robust hue constraints. Nevertheless, a few points in MAXI J1348--630 deviate from the \textit{RXTE}-defined regions, possibly because of averaged variable power spectra and the presence of a type-A QPO.
The trajectories of MAXI J1348--630 and MAXI J1820+070 exhibited roughly consistent patterns over different energy bands, whereas Swift J1727.8--1613 was an exception during its very high state, caused by an additional low-frequency component in the low energy band.
We found that the very high state can be identified through the power color-color diagram, exhibiting a hue similar to that of the hard intermediate state but not forming a loop pattern.
We also investigated the relationship between hue and hardness, and find that although they are generally anti-correlated, they provide consistent timing for the spectral state transitions.
%In addition, in contrast to \citet{lucchini2023variability}, we did not observe a clear lead of the hue evolution relative to the hardness evolution.

% %We present an analysis of four black hole low-mass X-ray binaries (BH LMXBs) observed with {\it Insight}-HXMT using the power-color method to investigate their fast X-ray variability. 
% We confirm that, in the low-energy (LE) band, the outburst evolution of all four sources follows the characteristic track previously reported by \citet{heil2015power}. 
% For three systems, the power-color tracks across different energy bands (LE, ME, and HE) are nearly identical in shape, showing only minor positional shifts. 
% Swift J1727.8--1613 represents an exception during its very high state (VHS), where the LE power spectra reveal an additional low-frequency component, likely associated with accretion disk instabilities. 
% After subtracting this component, the LE track aligns closely with the ME and HE tracks. 
% Our results suggest that the inward motion of the accretion disk during the VHS enhances high-frequency variability, causing the tracks to migrate toward the lower-right region of the power color-color diagram instead of forming a closed loop.
\end{abstract}


\section{INTRODUCTION} \label{sec:introduction}
In our Galaxy, a large population of low mass X-ray binaries (LMXBs) has been discovered \citep[e.g.,][]{liu2007catalogue,Avakyan2023}.
They typically consist of a compact object (either a neutron star or a black hole) and a companion star.
The compact object accretes material from the companion star, and the accreted matter gradually spirals inward owing to the viscosity, releasing gravitational potential energy in X-rays.
X-ray spectra of black hole (BH) LMXBs are generally composed of two primary components: a multi-temperature blackbody originating from a geometrically thin, optically thick accretion disk, and a non-thermal component produced by Compton scatterings in a geometrically thick, optically thin hot corona and/or jets \citep[e.g.,][]{Done2007}. 

Since most BH LMXBs are transient, they usually undergo a series of canonical states during outbursts according to their spectral and timing properties \citep[][]{remillard2006x,Ingram2019}. 
%A complete evolution is briefly summarized as follows.
At the beginning of an outburst, a source transits from the {\it quiescence state} to the \textit{hard state} (HS), where the emission is dominated by a powerlaw-like component with a photon index of 1.5--2.1. 
In this state, the power density spectrum (PDS) typically exhibits a strong, flat-topped, and broadband noise.
Along with the evolution, the system then enters into the \textit{hard intermediate state} (HIMS), characterized by a gradual softening of the energy spectrum and the presence of type-C quasi-periodic oscillations (QPOs) in the PDS \citep{done2002accretion}. 
Subsequently, the system may move to the \textit{soft intermediate state} (SIMS), marked by a prominent Type-B QPO at $\sim$6\,Hz in the PDS, without significant spectral variations \citep{belloni2009states}. 
At a higher accretion rate, the source evolves into the \textit{soft state} (SS), where the emission is dominated by the thermal disk as its inner radius moves inward. 
The PDS in this state remains to be broad but with a very low amplitude.
%
%Finally, the system returns to the hard and quiescence states.
In addition, some sources present a peculiar state, known as the \textit{very high state} (VHS) or the \textit{steep power-law state} (SPL) in the literature. 
This state is characterized by a steeper power-law spectrum with a photon index of $\Gamma > 2.4$, and probably presents a hard X-ray tail extending to the gamma-ray band \citep{mcclintock2003black}. 
The PDS in this state typically exhibits strong low-frequency QPOs \citep{mcclintock2003black}. 
Although originally identified during the brightest phases of outbursts \citep{tanaka1995x}---suggesting that the accretion disk may be even closer to the black hole than that in the soft state---subsequent observations have revealed the VHS in less luminous phases as well. 
% Therefore, the SPL state is now more commonly defined by its steep spectral slope rather than its luminosity \citep{mcclintock2003black}.

The outburst evolution of X-ray binaries has been conventionally studied using hardness--intensity diagrams (HIDs), in which the trajectory exhibits a characteristic ``q''-shaped hysteresis loop \citep{fender2005unified}. 
It provides an effective approach for tracking spectral state transitions while no temporal properties are taken into account.
On the other hand, the root mean square (RMS)--intensity diagram was proposed to describe outbursts by considering the intensity of overall variabilities \citep{munoz2011fast}, which however still lacks the detailed structure of PDS.
%To incorporate temporal features, root mean square (RMS) intensity diagrams were introduced later.  
%The RMS-intensity diagrams are useful to trace the evolution of the overall variability strength across different spectral states.
%However, they still lack information about the detailed shape of the PDS\citep{munoz2011fast}.
To characterize the evolution of PDS, \citet{heil2015power} proposed a novel method by introducing the concept of \textit{power-colors}---ratios of integrated power within different frequency bands in PDS. 
They found that by combining two power colors into a two-dimensional diagram, this method provides an effective means to classify outburst states.
Previous investigations have mainly been conducted using {\it Rossi X-ray Timing Explorer} ({\it RXTE}) data in the energy band of 2--13 \,keV.
We note that the temporal properties of XRBs are typically energy-dependent \citep{gierlinski2005patterns}. 
This motivates us to study the power-colors in different energy bands.
In this study, we conduct a detailed analysis to examine multiband power-color evolutions in three BH LMXBs (MAXI J1348--630, MAXI J1820+070, and Swift J1727.8--1613) using the broadband satellite Insight-Hard X-ray Modulation Telescope ({\it Insight}-HXMT).
These three BH LMXBs were selected owing to the availability of very high cadence observations covering their giant outbursts.

\section{METHOD AND DATA REDUCTION} \label{sec:METHOD AND DATA REDUCTION}

\subsection{The ``Power-Colour" methodology}\label{subsec:methodoloy}
The power color-color diagram (PCCD) was first introduced by \citet{heil2015power}, and here we briefly describe this methodology.
%utilized data from the \textit{Rossi X-ray Timing Explorer} (RXTE) to analyze outbursts from 12 BH LMXBs and all observations of Cygnus X-1 and Aql X-1. 
%For each observation, light curves were conducted in the 2--13 keV energy band, with each observation ID corresponding to several hours of exposure. 
%Power spectra were extracted from RXTE PCA (Proportional Counter Array) lightcurves binned up to 1/128 s with 512 s segment size such that the lowest frequency was 1/512 s.
In their study, for a given observation, the power spectrum was extracted in the frequency range of 1/512--64\,Hz, and the considered energy band for the power spectrum was 2--13 \,keV.
% In our study, we extend the energy range to higher energies. 
Following previous studies, two power colors are defined to characterize the power-spectral shape, which are the ratios of integrated Poisson-noise-subtracted power (i.e., the net variance) over different Fourier frequency ranges.
The first power-color (PC1) indicates the variance in 0.25--2.0\,Hz divided by that in 0.0039--0.031\,Hz, and the second power-color (PC2) is the variance in 0.031--0.25\,Hz divided by that in 2.0--16.0\,Hz.
These frequency intervals are chosen to be contiguous and geometrically spaced, such that they are evenly separated in logarithmic frequency space.
The power color--color diagram is constructed in logarithmic space, with $\log_{10}(\mathrm{PC1})$ and $\log_{10}(\mathrm{PC2})$ on the axes.
Following \citet{heil2015power}, a reference point located at $(\mathrm{PC1}, \mathrm{PC2}) = (4.51920,\,0.453724)$ in linear space is adopted, and its logarithmic coordinates define the origin for measuring angles in the diagram.
A line with a slope of $-1$ in the $\log$--$\log$ plane passing through this reference point is defined as the zero-angle axis.
Each observation is represented as a point in the $\log_{10}(\mathrm{PC1})$--$\log_{10}(\mathrm{PC2})$ diagram, and the angle between the line connecting this point to the reference point and the zero-angle axis, measured clockwise in logarithmic space, is defined as the \textit{hue}, by analogy with a color wheel.

This provides a simple and model-independent method for describing the power spectrum shape.
Furthermore, they found that different outburst states correspond to different hues in the power color-color diagram (Table~2 and  Figure~2 in \citet{heil2015power}).
Thus, the trajectory in the power color-color diagram is a diagnostic tool for tracking the state evolution during outbursts.

% Based on the trajectories, \citet{heil2015power} found that all sources follow a consistent evolutionary pattern during the rising phase of their outbursts: starting in the HS, transitioning through the HIMS and SIMS, and eventually reaching the soft state.
% This progression corresponds to a clockwise path in the diagram.And then the tacjectories from the SS back to the HS with a anti-clockwise path.
% Therefore, they simply based on their experience divided power colour-colour diagram into four regions:HS region, HIMS region, SIMS region, SS region,as we can see in Figure~\ref{fig:angle}.It can describe BH LMXBs state transitions and contains temporal information. 

%As shown in Figure~\ref{fig:angle}, different spectral states correspond to different ranges of hue: the HS spans from $340^\circ$ to $140^\circ$, the HIMS from $140^\circ$ to $220^\circ$, the SIMS from $220^\circ$ to $330^\circ$, and the SS from $300^\circ$ to $20^\circ$.

% \begin{figure}[htp]
%     \centering
%     \includegraphics[width=0.95\columnwidth]{figure/pc.pdf}
%     \caption{\textbf{The locations of various states in the power color-color diagram, where the purple area in the top left corner presents both the hard and soft states. The colored background shows the {\it RXTE}-defined hue ranges adopted from \citet{heil2015power}.}}
%     \label{fig:angle}
% \end{figure}

\begin{figure}[htbp!]
    \centering

    \begin{subfigure}[t]{0.4\textwidth}
        \centering
        \includegraphics[width=\linewidth]{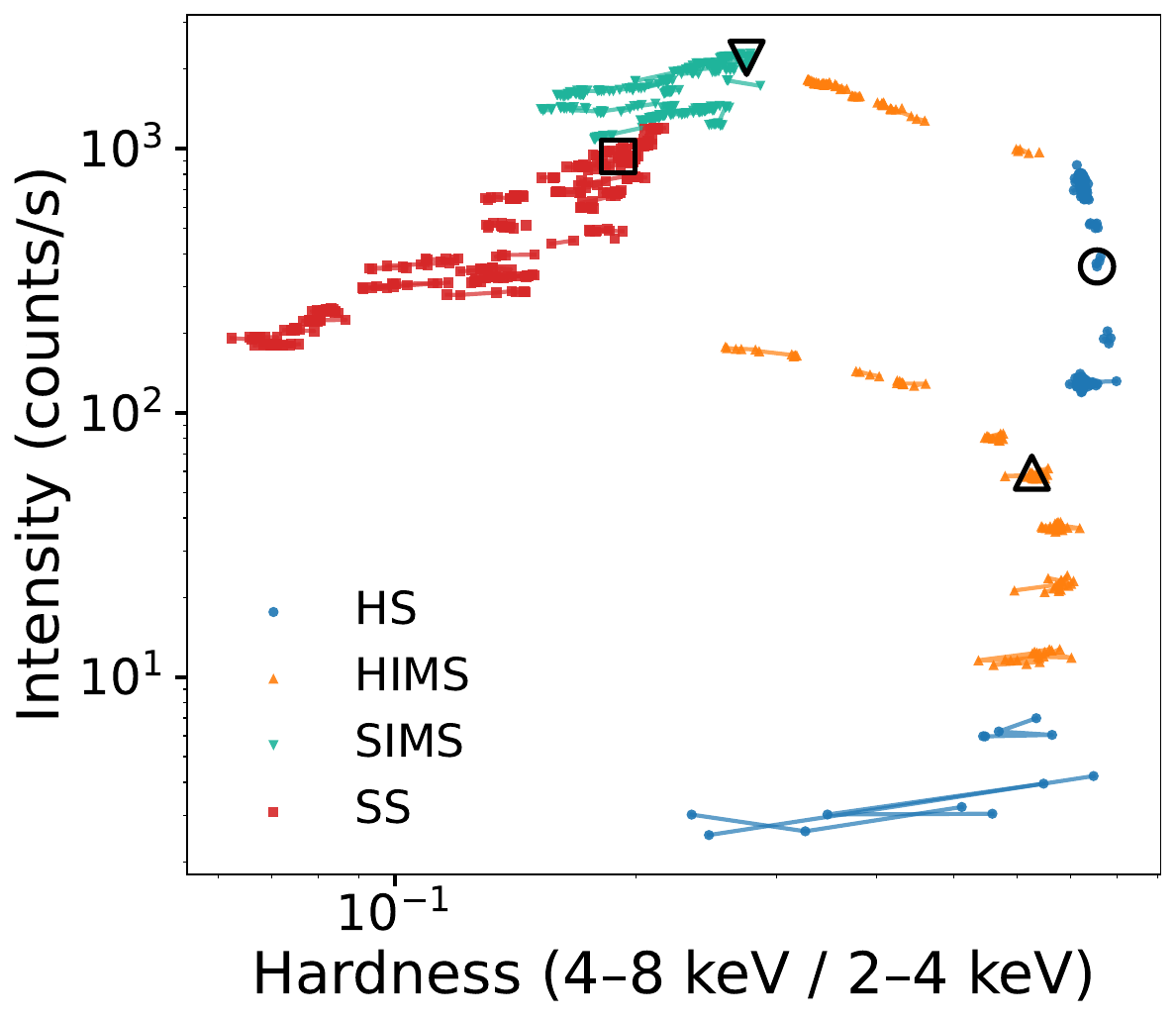}
        \caption{MAXI J1348--630}
    \end{subfigure}
    \hfill
    \begin{subfigure}[t]{0.4\textwidth}
        \centering
        \includegraphics[width=\linewidth]{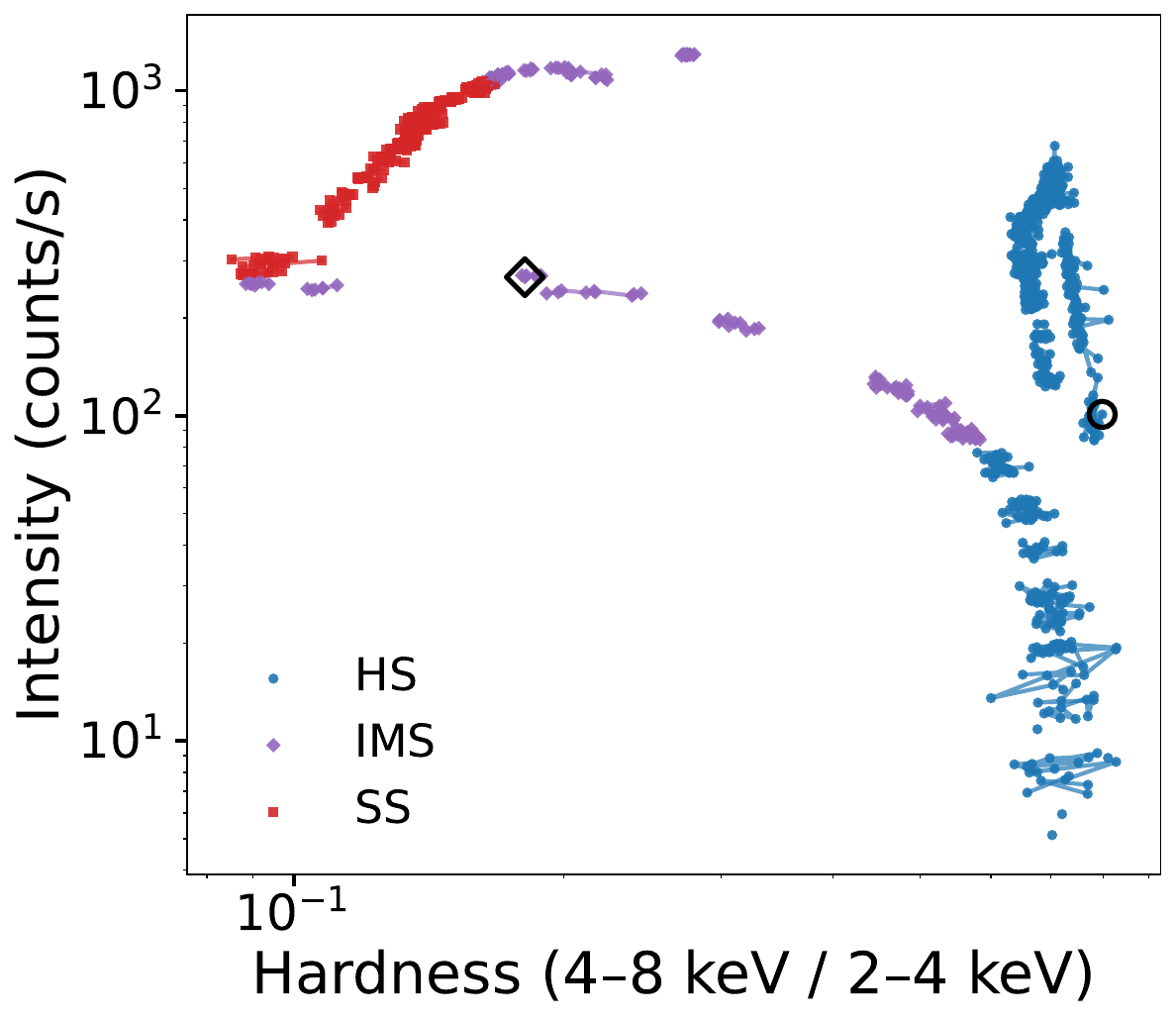}
        \caption{MAXI J1820+070}
    \end{subfigure}
    
    \begin{subfigure}[t]{0.4\textwidth}
        \centering
        \includegraphics[width=\linewidth]{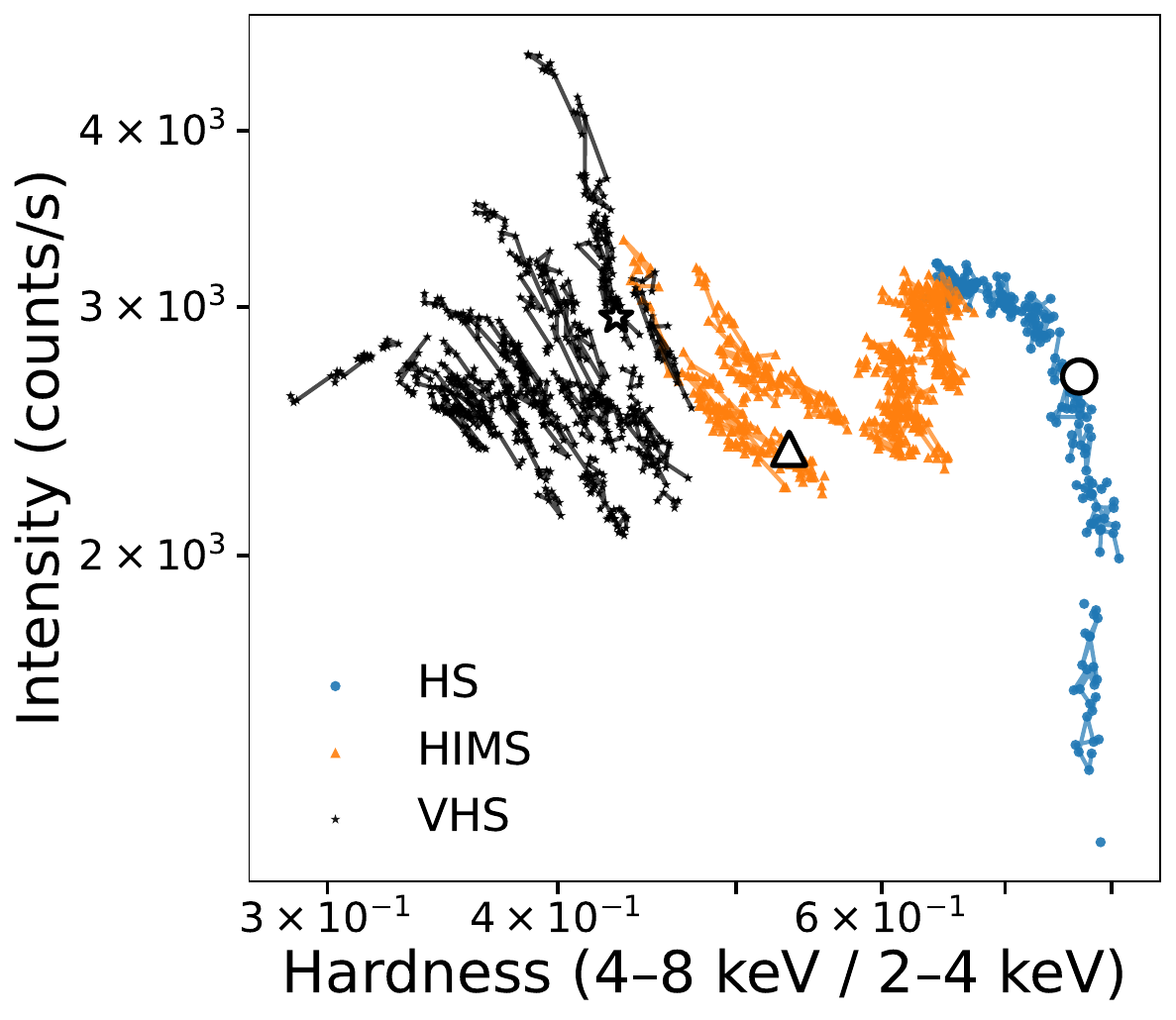}
        \caption{Swift J1727.8--1613}
    \end{subfigure}

    \caption{HIDs of three black hole X-ray binaries, where the hardness is defined as the count ratio between 4-8\,keV and 2-4\,keV. 
    Each point corresponds to a time segment of 512\,s. 
    Different colors and markers indicate different outburst states (see text): HS (blue circles), HIMS (orange upward triangles), SIMS (cyan downward triangles), IMS (purple diamonds), SS (red squares), and VHS (black stars). 
    Large empty symbols indicate representative observations whose PDSs are shown in Figure~\ref{fig:allpds}.
    }
    \label{fig:3HID}
\end{figure}

\begin{figure*}
\centering
\includegraphics[width=\textwidth]{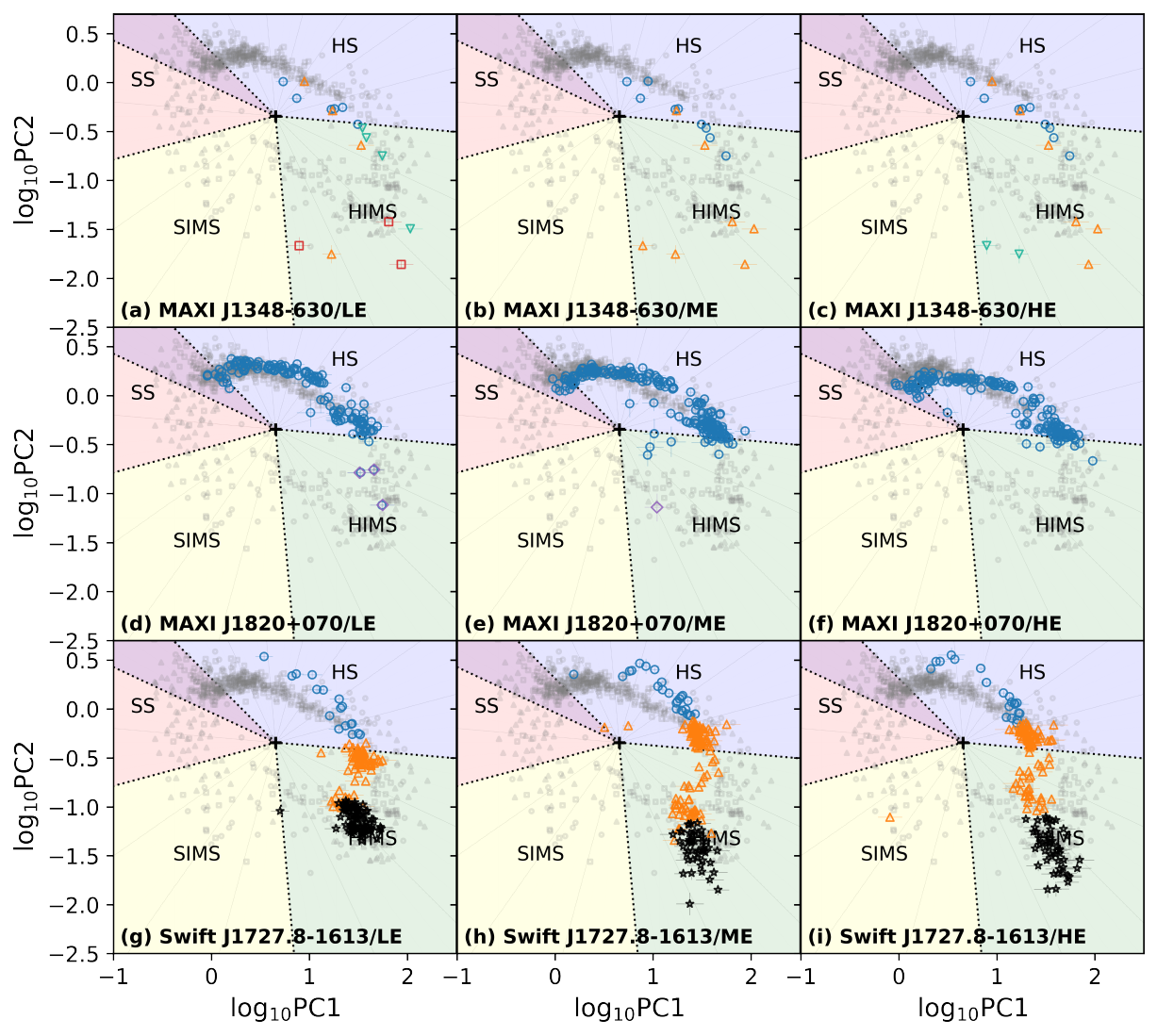}
\caption{
Power color-color diagrams of three black hole X-ray binaries in different energy bands.
Each point represents a ``good" observation.
Different colors and markers indicate different outburst states, i.e.,
HS (blue circles), HIMS (orange upward triangles),
SIMS (cyan downward triangles), IMS (purple diamonds), SS (red squares), and VHS (black stars).
For comparison, three {\it RXTE}-observed black holes are included, i.e., GX 339--4 (grey squares), H1743--322 (grey triangles), and XTE J1550--564 (grey circles).
These points are adopted from \citet{Gardenier_2018}.
The colored background shows the {\it RXTE}-defined hue ranges adopted from \citet{heil2015power}.}
% \textcolor{blue}{Don’t put all sentences and paragraphs in bold; only mark the specific words or sentences that have been changed. Otherwise, marking everything is the same as not marking anything at all!}

\label{fig:9pc}
\end{figure*}

%They find that the different objects follow a distinct and consistent track in a power-‘color-color’ diagram and show that this track lends itself to a single-value parameterisation of the power-spectral shape (and thus source state), which they call the ‘hue’ of the power spectrum.
%In the following analysis, we apply this method to data from the \textit{Insight}-HXMT satellite to test the validity of these conclusions. 
%In addition, we extend the investigation to higher energy bands to explore whether the same behavior persists beyond the RXTE energy range.

\subsection{Data reduction} \label{subsec:data reduction}

We used the data from the {\it Insight}-HXMT, which is equipped with three payloads covering different energy ranges: the Low Energy X-ray Telescope (LE) in 1--15 \,keV, the Medium Energy X-ray Telescope (ME) in 5--30 \,keV, and the High Energy X-ray Telescope (HE) in 20--250 \,keV \citep{Chen2020LE,Cao2020ME,Liu2020HE}.
%The broad energy coverage provided by \textit{Insight}-HXMT is well-suited to meet the requirements of our study.
%In particular, the HE detector offers an exceptionally large effective area in the hard X-ray band ($>$20 keV), significantly enhancing its capability to detect faint sources and capture rapid transient phenomena in this energy regime.
%Furthermore, \textit{Insight}-HXMT has high temporal resolution, reaching up to 25 microseconds, making it highly suitable for investigating rapid variability phenomena in neutron star and black hole systems, such as QPOs and X-ray bursts.
We conducted the data analysis using the {\it Insight}-HXMT Data Analysis Software (HXMTDAS, version 2.06) with the latest calibration database (CALDB, version 2.07).
We only considered data from small field-of-view (FOV) detectors.
The good time intervals (GTIs) were selected according to the following criteria: pointing offset angle $<$ 0.04$^\circ$, elevation angle (ELV) $>$ 10$^\circ$, geomagnetic cutoff rigidity (COR) $>$ 8\,GV, and at least 300\,s away from the South Atlantic Anomaly (SAA).
For the LE, an additional constraint of $\mathrm{DYE\_ELV} > 30^\circ$ was applied to prevent optical contamination from the bright Earth.
% To optimize the signal-to-noise ratio, in the following analysis, we adopted energy bands of 2--8 \,keV, 10--30 \,keV, and 30--80 \,keV for LE, ME and HE, respectively.

%The joint spectral fitting of all three instruments is performed using \texttt{XSPEC} (version 12.12.1), with a systematic uncertainty of 0.5\% included in the modeling.
% \begin{table*}
% \centering
% \caption{Summary of {\it Insight}-HXMT observations. 
% For each source, the number of observations, total exposure, and time span for different energy bands are listed.
% }
% \caption{Summary of {\it Insight}-HXMT observations.
% For each source, the number of observations and total exposure in different energy bands are listed.
% The time span corresponds to the overall HXMT observational coverage of each source, as described in the text.}
% \label{tab:obs_summary}
% \begin{tabular}{lccccccc}
% \hline\hline
% Source 
% & \multicolumn{2}{c}{LE} 
% & \multicolumn{2}{c}{ME} 
% & \multicolumn{2}{c}{HE}
% & Time span (MJD) \\
% \cline{2-3} \cline{4-5} \cline{6-7}
% & $N_{\rm obs}$ & Exposure (ks)
% & $N_{\rm obs}$ & Exposure (ks)
% & $N_{\rm obs}$ & Exposure (ks)
% & \\
% \hline
% MAXI~J1348$-$630 
% & 202 & 267.9 
% & 219 & 628.9 
% & 216 & 473.4
% & 58510--58652 \\

% MAXI~J1820$+$070 
% & 320 & 645.2 
% & 330 & 1073.8 
% & 320 & 782.6
% & 58191--58412 \\

% Swift~J1727.8$-$1613 
% & 291 & 477.8 
% & 296 & 717.4 
% & 286 & 557.6
% & 60181--60222 \\
% \hline
% \end{tabular}
% \end{table*}

\begin{table*}
\centering
\caption{
Summary of {\it Insight}-HXMT observations. 
For each source and energy band, the total number of observations, the number of selected ``good" observations (see text), and the corresponding exposure time are listed.
%satisfying the $3\sigma$ criterion in all four frequency bands (including combined observations), and the total exposure time of ``good'' observations are listed.
}
\label{tab:good_obs}
\begin{tabular}{lcccc}
\hline\hline
Source & Band & $N_{\rm obs}$ & $N_{\rm good}$ & Exposure (ks)  \\
\hline

MAXI~J1348$-$630 
& LE & 202 & 17 & 93.70  \\
& ME & 219 & 17 & 131.04  \\
& HE & 216 & 18 & 139.22  \\

\hline

MAXI~J1820$+$070 
& LE & 320 & 150 & 398.98  \\
& ME & 330 & 200 & 707.48 \\
& HE & 320 & 167 & 512.37  \\

\hline

Swift~J1727.8$-$1613 
& LE & 291 & 176 & 363.03 \\
& ME & 296 & 178 & 496.37 \\
& HE & 286 & 158 & 409.60  \\

\hline
\end{tabular}
\end{table*}

In this study,  the observations span MJD~58510--58652 for MAXI~J1348$-$630, 
MJD~58191--58412 for MAXI~J1820$+$070, 
and MJD~60181--60222 for Swift~J1727.8$-$1613.
Detailed information on the observations is summarized in Table~\ref{tab:good_obs}.
% \textbf{Only observations for which the variance was constrained at a significance level higher than $3\sigma$ in all four frequency bands were used to calculate the power colors. 
% Following \citet{heil2015power}, such observations are referred to as 
% ``good'' observations.
% Because the count rate is relatively low in the soft state, and the effective 
% area of {\it Insight}-HXMT is smaller than that of {\it RXTE}, 
% some observations do not satisfy the above requirement individually. 
% To improve the signal-to-noise ratio, the observations of MAXI J1348--630 in all three energy bands were combined according to their positions in the HID (each group fell within a 5\% range in hardness and a 10\% range in intensity, with a maximum time span of five days. Different grouping schemes were tested and found not to significantly affect the results).
% Detailed information on the total observations, good observations, 
% and the corresponding exposure times for each source is summarized 
% in Table~\ref{tab:good_obs}.}
%
%\textbf{For each ``good'' observation, light curves were extracted separately from the LE, ME, and HE instruments in the energy ranges of 2--8\,keV, 10--30\,keV, and 30--80\,keV, respectively, with a time resolution of 1/128\,s.}
For each observation, we extracted lightcurves in the energy ranges of 2--8 \,keV, 10-30 \,keV and 30-80 \,keV with a time resolution of 1/128 \,s. 
We calculated PDSs from the lightcurves using the \texttt{stingray}\footnote{\url{https://docs.stingray.science/en/stable/index.html}} Python package with a time segment of 512\,s, corresponding to the considered frequency range of 1/512-64\,Hz.
The PDSs were normalized using the Leahy normalization, which facilitates the subtraction of the expected Poisson white noise level with a value of 2 \citep{Leahy1983OnSF}.
The power colors were calculated from PDSs by integrating the noise-subtracted powers over predefined frequency bands, following \citet{heil2015power}, and the uncertainties were estimated by using the error propagation.
Because the effective area of {\it Insight}-HXMT is smaller than that of {\it RXTE}, the resulting large uncertainties may significantly affect the positions in the PCCD, particularly for the soft state when the variability is weak.
To minimize this effect, following \citet{heil2015power}, only observations for which the variance is detected at a significance level above $3\sigma$, i.e., the ratio of the integrated power to its uncertainty exceeds 3, in all frequency bands are selected. 
These observations are referred to as ``good" observations and are used in the subsequent analysis.
Since most observations of MAXI J1348--630, as well as the intermediate state and soft state observations of MAXI J1820+070, do not satisfy the above criterion, we therefore combined them according to their positions in the HID, with each group spanning no more than 5\% in hardness and 10\% in intensity and within a time windows of five days. 
Different grouping schemes were tested and found little influence on the results.
However, the hard state observations of MAXI J1820+070
and most observations of Swift J1727.8--1613 satisfy the $3\sigma$ criterion individually owing to their higher signal-to-noise ratios, and therefore no observation combination was required.
% {(JL: but how about the case for other sources? for example, the soft state for maxi j1820?}

%
In addition, we also studied an additional low frequency component (see Section~\ref{subsec:1727}), which appears only in the LE band, in the VHS of Swift J1727.8--1613.
To check its influence, we fitted the PDSs with multiple Lorentzian components, and estimated power colors by excluding the contribution of this low frequency component based on the best-fitting model.
In practice, we converted the PDSs into pseudo-energy spectra using the \texttt{ftflx2xsp}\footnote{\url{https://heasarc.gsfc.nasa.gov/docs/software/lheasoft/help/flx2xsp.html}} tool following \citet{Ingram2012}, and performed the fitting with {\sc XSPEC}.
All uncertainties quoted in this study correspond to a confidence level of 68\%.

\begin{figure*}
\centering
\includegraphics[width=\textwidth]{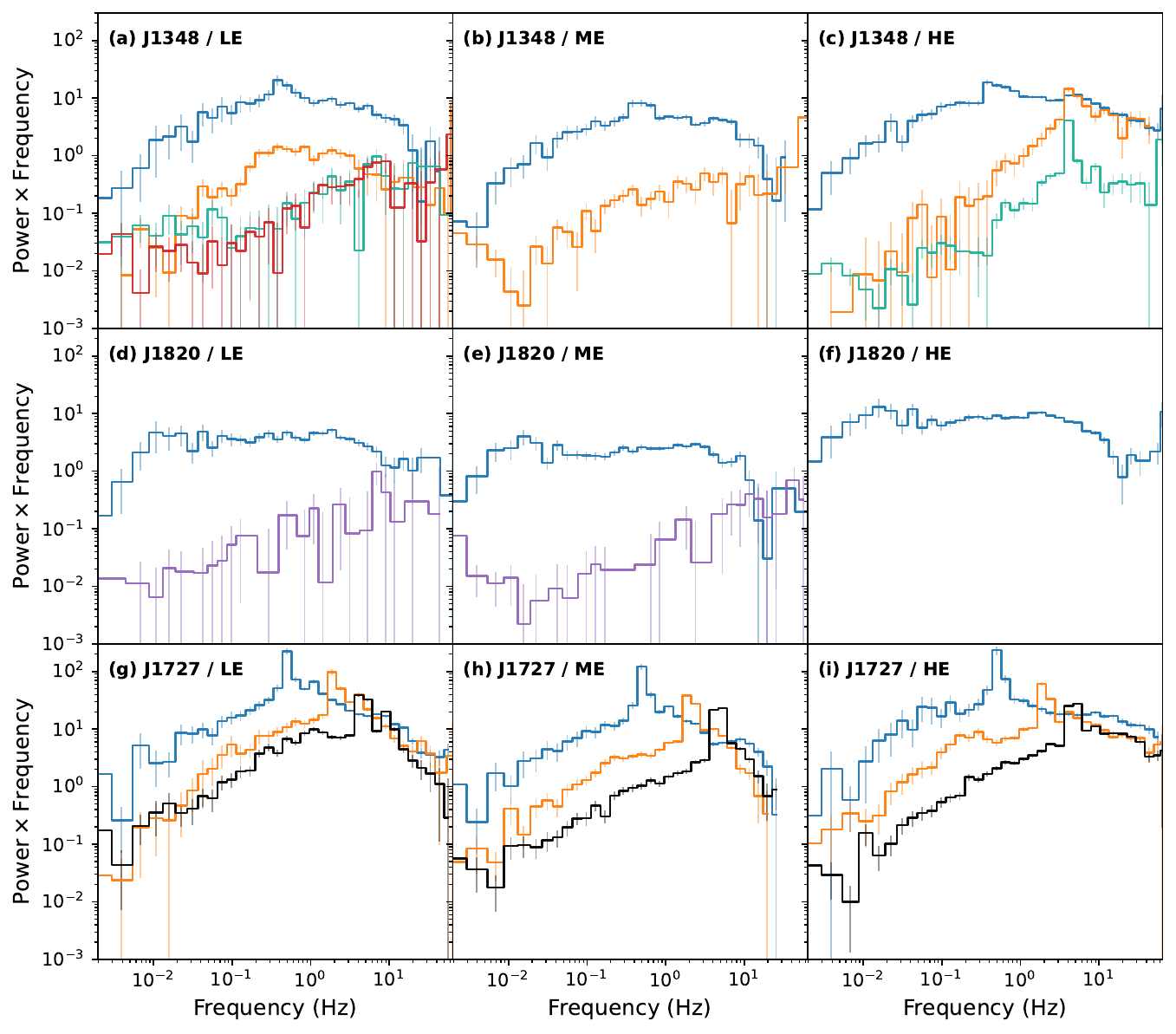}
\caption{
Representative PDSs from selected ``good" observations (marked in Figure~\ref{fig:3HID}) corresponding to different outburst states (HS in blue, HIMS in orange, SIMS in cyan, IMS in purple, SS in red, and VHS in black) of MAXI J1348--630 (J1348), MAXI J1820+070 (J1820) and Swift J1727.8--1613 (J1727). 
The Poisson noise has been subtracted. 
We note that for some soft and intermediate states, no PDS satisfies the selection criterion, i.e., the ratios of integrated powers to their uncertainties exceed 3 in all frequency bands.
%Not all spectral states contain ``good'' observations; therefore, no representative PDSs are shown for those states.
}
% \caption{
% Averaged PDSs for different outburst states (HS in blue, HIMS in orange, SIMS in cyan, IMS in purple, SS in red, and VHS in black) of MAXI J1348--630 (J1348), MAXI J1820+070 (J1820) and Swift J1727.8--1613 (J1727).
% The Poisson noise has been subtracted.}
%Different colors indicate different spectral states: .
%combined data. Each row corresponds to a source and each column to one energy band (LE, ME, and HE from left to right). The sources are MAXI J1348--630 (J1348), MAXI J1820+070 (J1820), and Swift J1727.8--1613 (J1727).
%All PDSs are logarithmically rebinned with a factor of 0.2, and the Poisson noise level has been subtracted.}
% 

\label{fig:allpds}
\end{figure*}

\begin{figure}
\centering
\includegraphics[width=0.95\columnwidth]{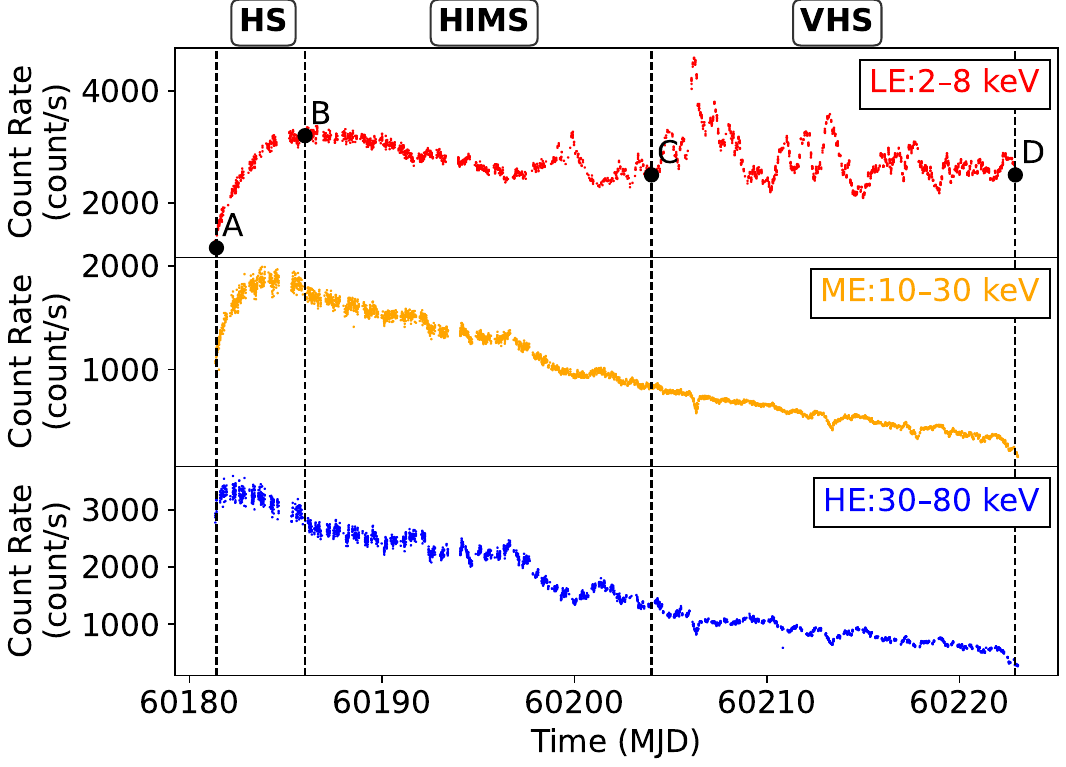}
\caption{
Long-term lightcurves of Swift J1727.8--1613 during its 2023 outburst in the energy bands of 2-8 \,keV (upper), 10-30 \,keV (middle) and 30-80 \,keV (bottom). 
Vertical dashed lines separate different outburst evolution states, as provided by previous studies \citep{liu2024broadbandxrayspectralproperties, cao2025spectral}.
%The panels from top to bottom correspond to the LE, ME, and HE bands, respectively. 
% Four vertical dashed black lines mark significant epochs: the first (MJD 60181) and fourth (MJD 60222) indicate the start and end of the observation period, while the second (MJD 60197) and third (MJD 60204) correspond to the beginning and end of the first flaring episode in the flare state. 
% The intersections of these vertical lines with the LE light curve are marked as points A, B, C, and D, respectively.
}
\label{fig:lc}
\end{figure}

\begin{figure*}
\centering
\includegraphics[width=\textwidth]{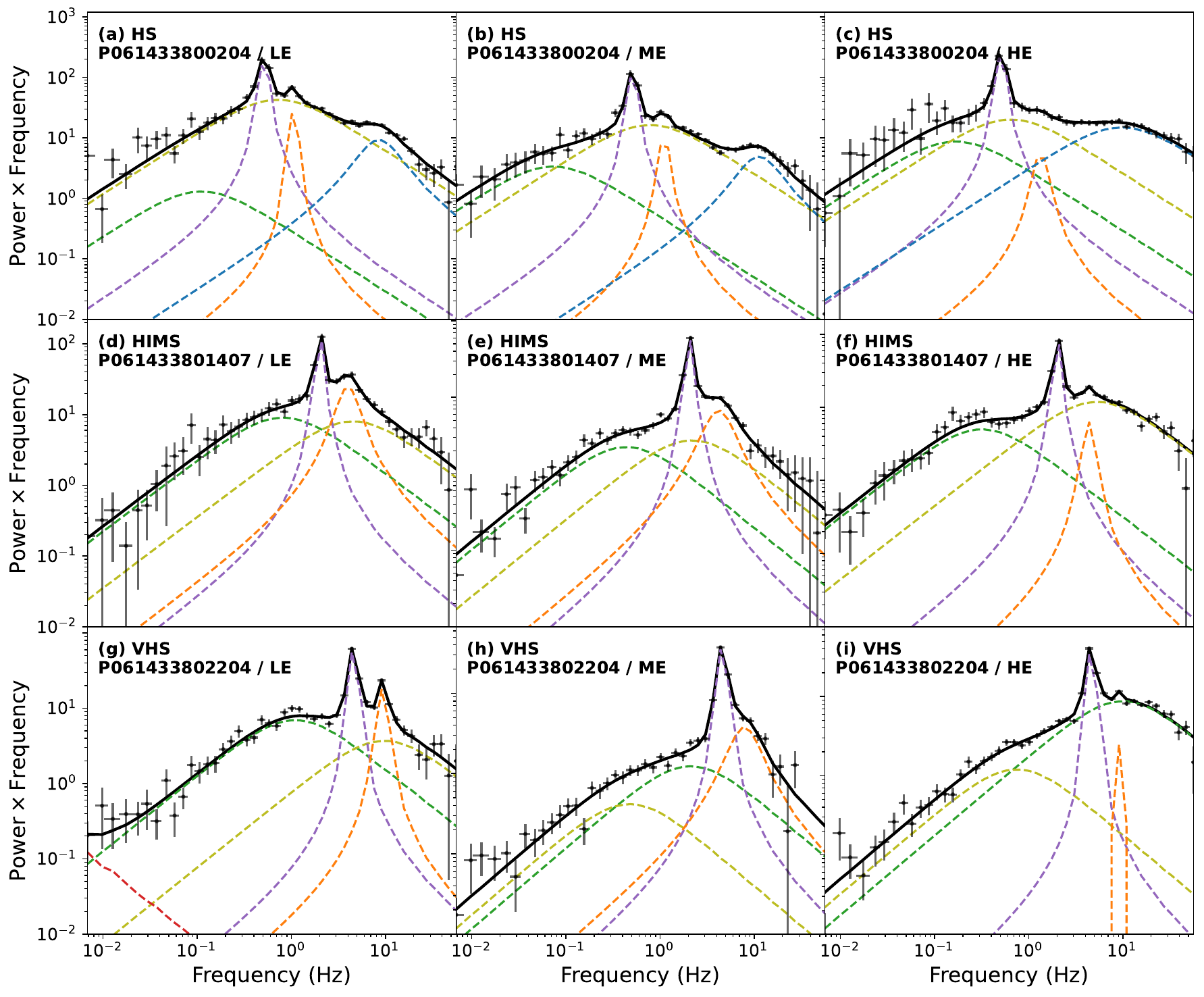}
\caption{
Representative PDSs of Swift J1727.8--1613 observed by {\it Insight}--HXMT in the LE (2--8 \,keV; left), ME (10--30 \,keV; middle), and HE (30--80 \,keV; right) energy bands at different spectral states.
The colored dashed curves indicate the Lorentzian components used to fit the QPOs and broadband noise.
Note that, in the VHS, an additional low frequency component (red line) is required only for the LE band.
}
\label{fig:9pds}
\end{figure*}

\section{RESULTS} \label{sec:results}

The hue ranges corresponding to various outburst states were determined from previous 2--13 \,keV {\it RXTE} observations. However, the power spectra in different energy bands are typically not identical. On the other hand, even when similar energy bands are considered, they may differ because of the discrepancy in the effective areas of different instruments. 
In this section, we present the power colors obtained from {\it Insight}-HXMT observations and compare them with the hue ranges determined by {\it RXTE} observations.

We first present the HIDs of the three sources in Figure~\ref{fig:3HID}, with different colors indicating different outburst states according to previous temporal and spectral studies (see below for details). 
The HID of MAXI J1348--630 displays a typical ``q"-shaped trajectory commonly observed in LMXBs.  
The ``q"-shape is also observed in MAXI J1820+070 but with an additional structure associated the preceding HS. 
For Swift J1727.8--1613, the HID is more complex, in particular for the flaring state as reported by \citet[e.g.,][]{yu2024timing}.
In general, these HIDs imply that there are various outburst states in our sample, and therefore these three sources are suitable for studying corresponding positions in the PCCDs.
%
%a flare state is also present, consistent with the results reported by \citet{yu2024timing}.
%
%In Figure~\ref{fig:3HID}, we find that the locations of the spectral states adopted from previous studies in the HID are generally consistent with those expected from the typical q-shaped evolution of LMXBs \citep{fender2005unified}. Therefore, we adopt the state classifications reported in the literature throughout this work.

%\textbf{Firstly, we applied the same S/N criterion as \citet{heil2015power} to {\it Insight}-HXMT observations, requiring that the variance is detected at a significance level greater than $3\sigma$ in all four frequency bands used to compute the power colours. However, we found that a large fraction of individual observations—particularly those in the soft and intermediate states of MAXI J1348–630 and MAXI J1820+070—do not satisfy this strict criterion, primarily due to the relatively low count rates compared to RXTE. Applying this requirement to individual observations would therefore exclude most of the data and prevent a meaningful application of the power colour method to HXMT observations.
% , we ignored this limitation which every point should satisfy variance detected at >3σ in all frequency bands
%To improve the S/N, we combined observations within the same spectral state for each source and each energy band, and computed an averaged power spectrum for each state, the power colours were then calculated from these PDSs using the direct integration method.}

Figure~\ref{fig:9pc} shows multi-band power color--color diagrams for MAXI J1348--630, MAXI J1820+070 and Swift J1727.8--1613, in which each point refers to one  ``good" observation.
Different colors and markers indicate various outburst states identified through detailed temporal and spectral studies.
%\textbf{To investigate the evolution of the PDS shapes in different states together with the power color, we selected representative PDS for each state in the three energy bands, as shown in Figures~\ref{fig:allpds}.}
For MAXI J1348--630, we adopted results from \citet{zhang2020nicer} and \citet{carotenuto2026rapid}, who classified the outburst into four states, i.e., HS (MJD~58509--58517 and 58604--58652), HIMS (MJD~58517--58522.6 and 58597--58604), SIMS (MJD~58522.6--58542) and SS (MJD~58542--58597).
As shown in Figures~\ref{fig:9pc} (a)-(c), LE points during the HS generally lie within the expected region of {\it RXTE}-defined hue angles, whereas for ME and HE bands some of the points are located outside this region.
For the HIMS, the points in all three energy bands are generally located within the {\it RXTE}-defined HIMS region, with only a few near the boundary between the {\it RXTE}-defined HS and HIMS regions. 
For the SIMS, only the LE and HE points are available even after combining observations and all of them lie within the {\it RXTE}-defined HIMS region.
Similarly, for the SS, most observations fail to meet the requirement that the integrated powers should be detected at a significance level of $>3\sigma$, even after combining observations. 
Only a few observations (MJD 58546--58553) yield valid measurements of power colors in the LE band, and these points fall within the {\it RXTE}-defined HIMS region.
Overall, the HS and HIMS power colors of MAXI J1348--630 obtained with {\it Insight}-HXMT in all three energy bands are broadly consistent with the corresponding regions defined by {\it RXTE}, whereas the locations of the SIMS and SS points deviate from the previously defined regions.
%The LE-band points are mainly located within the {\it RXTE}-defined HIMS region, while the HE-band points lie close to the boundary between the {\it RXTE}-defined HIMS and SIMS regions.
%Overall, the HS and intermediate-state power colors obtained with {\it Insight}-HXMT in all three energy bands are broadly consistent with the corresponding regions defined by {\it RXTE}. 
%In contrast, for the SS, only the LE band data remain after combining observations, and these points are mainly located within the {\it RXTE}-defined HIMS region. 
%
To better investigate the evolution of PDS shapes, representative observations from different outburst states are shown in Figure~\ref{fig:allpds}, and their positions in the HIDs are marked in Figure~\ref{fig:3HID}.

For MAXI J1820+070, we adopted the state classification from \citet{peng2023possible}, who defined the HS as the periods of MJD~58190--58305 and 58390--58412, and the SS between MJD~58315 and 58380.
Because this source experienced rapid state transitions, they defined the periods of MJD~58305--58315 and 58380--58390 as the intermediate state without further subdivision.
As shown in Figures~\ref{fig:9pc} (d)–(f), the points of all three energy bands in the HS are located within the {\it RXTE}-defined region or close to its boundaries. 
Only a few IMS points satisfy the $3\sigma$ criterion, all of which are located within the {\it RXTE}-defined HIMS region. 
For the SS, the variability is so weak that even after combining observations, no reliable power colors can be obtained.
%the data still do not satisfy the $3\sigma$ criterion. 
%Therefore, it is not possible to determine whether the {\it Insight}-HXMT results are consistent with the corresponding {\it RXTE} results in this state.

% The intermediate-state points are distributed within the {\it RXTE}-defined HIMS and SIMS regions, and the state averaged points of different energy bands are all located in the {\it RXTE}-defined HIMS region.
% For the SS points, similar to the case of MAXI J1348--630, they are broadly scattered. 
% However, even when all available data is combined, errors remain significantly large due to its weak variability in this state, preventing us from drawing conclusive conclusions.
%The combined SS points lie within or close to the {\it RXTE}-defined SS region; however, their uncertainties are relatively large.This is likely due to the low photon counts rate in the soft state of MAXI J1820+070. In fact, the soft-state PDS of this source is the only case in this work where, even after combining observations, the variability does not reach a significance level above $3\sigma$ simultaneously in all four frequency bands. 

The outburst of Swift J1727.8--1613 in 2023 is peculiar, of which the lightcurves are shown in Figure~\ref{fig:lc}.
According to spectral variability and the detection of type-C QPOs, it was found in the HS until $\sim$MJD 60186, and was classified as in the HIMS between MJD 60186 and 60204 \citep{liu2024broadbandxrayspectralproperties, cao2025spectral}.
This source underwent its first flare between MJD 60197 and 60204, and the end of the first flare was considered as the transition from the HIMS to the VHS \citep{cao2025spectral}.
%In this paper, we arbitrarily identify this period as the HIMS.
After MJD 60204, this source exhibited more flares, which were considered in the VHS based on spectral studies\footnote{Also see discussions about the state classification using quasi-periodic oscillations \citep{Xu2025}.} \citep{cao2025spectral}.
The power colors of Swift J1727.8--1613 are presented in Figures~\ref{fig:9pc} (g)-(i).
{\it Insight}-HXMT's HS points are predominantly located within the {\it RXTE}-defined HS region, while those points associated with the HIMS are distributed in both HS and HIMS {\it RXTE} regions.
The position of the VHS in the power color-color diagram has not been previously studied, We found that they have hue angles similar to those of the HIMS.
We note that, unlike the evolution from the HIMS to the SS which presents a clockwise loop, the transition from the HIMS to the VHS corresponds to unchanged hue angles, but may be accompanied by a change in the values of PC1 and PC2.
%In addition, the VHS is unique in that its power colors are highly energy-dependent, showing a significant discrepancy between the LE and ME/HE bands.
In addition, after entering the VHS, the LE-band points show a relatively concentrated distribution in the PCCD, whereas ME and HE points become more dispersed and extend toward the lower-right region of the diagram.

\section{Discussion}

\subsection{Shifted hue angles in the SIMS and SS of MAXI J1348--630}
The state-dependent locations in the multiband PCCDs obtained with \textit{Insight}-HXMT are broadly consistent with those found with \textit{RXTE}, except for several MAXI J1348--630 observations classified as SIMS or SS that lie in the region usually associated with the HIMS.
These points should be considered with caution.
Because the state classification adopted here is based mainly on timing properties, with the detection of a type-B QPO taken as a characteristic signature of the SIMS \citep{zhang2020nicer}.
However, the timing evolution of MAXI J1348--630 during its state transition was not smooth.
As reported by \citet{Zhang2021}, the type-B QPOs were transient within individual observations.
In addition, there were rapid transitions between type-B and type-C QPOs \citep{Liu2022}, and their coexistence and interplay has been reported by \citet{Wang2026}.
In our analysis, since several observations had to be combined to accumulate the counting statistics, the resulting averaged power spectra may contain a mixture of different timing modes, shifting the power colors toward the nominal HIMS region.

In the soft state, a strong radio re-brightening and an increase in the X-ray rms variability were reported between MJD 58570 and 58590 \citep{carotenuto2026rapid}.
However, the statistics of our data during this period were insufficient to obtain reliable hue measurements.
In the SS, reliable power colors could be measured only in the LE band near the beginning of the SS (MJD 58546--58553).
This time window overlaps with the interval in which a weak and broad type-A QPO was detected \citep{zhang2023type}.
We therefore speculate that the type-A QPO may influence the measured power colors and shift the hue angles, although our current data do not allow testing this possibility conclusively.

 \subsection{A detailed investigation for Swift J1727.8-1613}\label{subsec:1727}
\begin{figure}
\centering
\includegraphics[width=0.95\columnwidth]{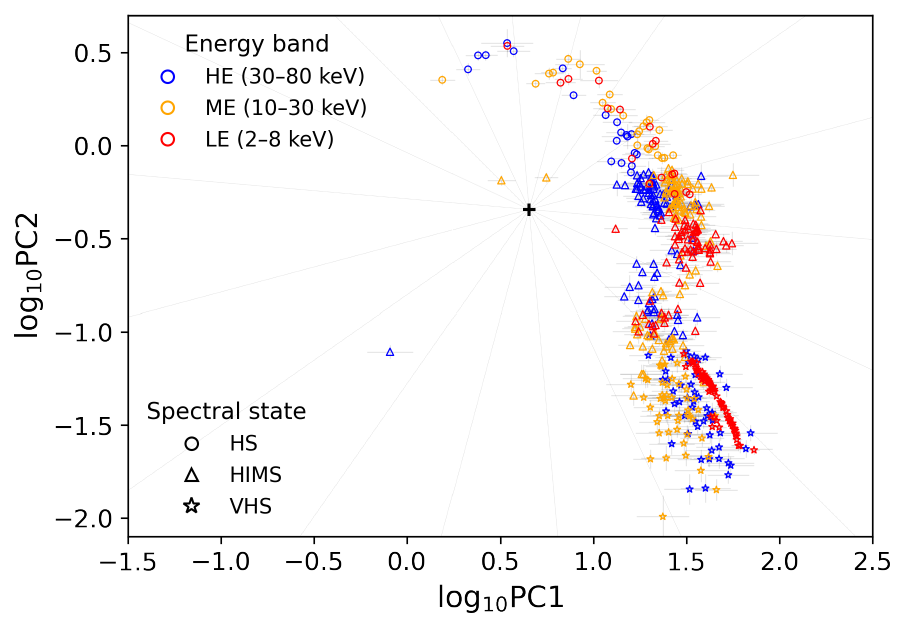}
\caption{
Power color-color diagrams of Swift J1727.8--1613, in which the low frequency excess detected in the LE band (see text) during the VHS has been subtracted.}
\label{fig:remove}
\end{figure}

\begin{figure}
\centering
\includegraphics[width=0.95\columnwidth]{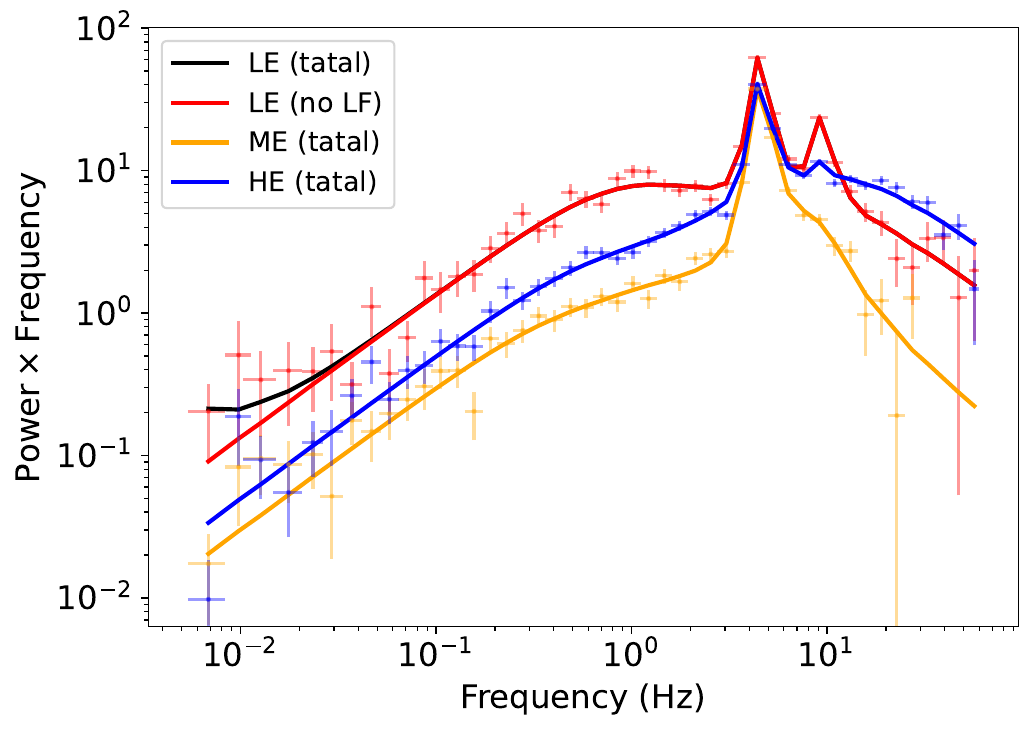}
\caption{
PDSs of three energy bands in the VHS of Swift J1727.8--1613.
The black, blue and yellow solid lines present the best-fitting models of LE, ME and HE bands.
The red line is the LE PDS after subtracting the additional low frequency component.
%The LE PDS is shown in black, and the corresponding PDS with the LF component removed is shown in red.
%The ME and HE bands are shown in yellow and blue, respectively.
}
\label{fig:4pds}
\end{figure}

We performed a more detailed analysis for Swift J1727.8--1613 to understand the peculiar behavior in its multi-band power color-color diagrams.
We first show long-term lightcurves of different energies in Figure~\ref{fig:lc} during its 2023 outburst, where ME and HE bands present a gradual decay, while the LE band exhibits flaring activities after MJD 60197.
This distinction implies a discrepancy in their corresponding power spectra.
As mentioned above, outburst evolution states at different epochs have been classified and identified according to temporal and spectral studies \citep[e.g.,][]{yu2024timing, cao2025spectral}.
We inspected their power spectral shapes and show representative examples in Figure~\ref{fig:9pds}.
Generally, the power spectra over different energy bands appear similar profiles, with the exception for the VHS where the LE band presents an additional enhancement at low frequencies, which may lead to the distinctive trajectory in the power color-color diagram.
To verify this hypothesis, we re-calculated LE's power colors by excluding this low frequency component.
In practice, we modeled the power spectra using a combination of several Lorentzian components for QPOs and broadband noise, and a constant accounting for the white noise. 
To obtain an acceptable fit, we employed two broadband zero-centered components and two QPO (fundamental and harmonic) components.
% An additional peaked noise around $\sim$10 Hz was included for the HIMS.
In addition, to account for the low-frequency excess in the VHS, we added an additional zero-centered Lorentzian component (red dashed line in the lower-left panel of Figure~\ref{fig:9pds}), and found that this component is required only in the LE band.
With the best-fitting parameters, we were able to compute the power colors after subtracting the contribution from this low frequency excess.
We found that, in this case, the LE trajectory becomes consistent with those of the ME and HE bands, extending toward the lower-right region of the RXTE-defined HIMS area in a similar manner, albeit with minor offsets (Figure~\ref{fig:remove}).
In Figure~\ref{fig:4pds} we present a direct comparison between PDSs of different energy bands, which shows that, after removing the low frequency component, the overall shape of the LE PDS becomes consistent with those of ME and HE.
%After subtracting this LF component from the LE PDS, the power colors were recomputed by integrating the residual power over the defined frequency bands.
%This is further supported by a direct comparison of the PDS, which shows that, after removing the LF component, the overall shape of the LE PDS becomes consistent with those of the ME and HE bands(Figure~\ref{fig:4pds}).

The spectra in the VHS have been studied by \cite{Xu2025} and \cite{He2025}.
They interpreted the soft X-ray flares as accretion rate fluctuations originating at and propagating from large radii.
The reason why the obtained Comptonized flux (e.g., hard X-rays) in this work is relatively stable might be due to the changes in the co-evolving covering fraction, which reflects the ratio of seed photons being Comptonized.
During flares the increasing in seed photons might be associated with the decreasing of the covering fraction, and thus these two effects cancel out, resulting in relatively stable Comptonized photons.
It is generally believed that changes of the covering fraction indicate the variations in the coronal geometry. However, the underlying physical processes remain unknown.
Our results suggest the presence of two different responses of the corona to the disk in the VHS:
One is caused by mechanisms similar to those of other states, leading to roughly identical power spectrum shapes across different energy bands.
The other is unique to the VHS, involving a certain negative feedback (such as the decreasing of the covering fraction) at low frequencies, which prevents variations in seed photons from affecting the high energy Comptonized flux.
%the number of

% \subsection{HIDs}

% \textbf{Figure~\ref{fig:3HID} shows the HIDs of the three sources. It can be seen that none of them exhibit the standard ``q-shaped'' HID evolution commonly observed in black hole X-ray binaries. 
% For MAXI J1348--630, the HID shows a shape more similar to a trapezoid. 
% In particular, the classification of the SS observations may not be fully 
% appropriate. The data points that turn toward higher hardness during the 
% soft state would traditionally be considered intermediate-state points 
% according to the canonical HID evolution. 
% Moreover, \citet{carotenuto2026rapid} reported a sudden radio re-brightening during the soft state, accompanied by a significant increase in the X-ray fractional rms. 
% They therefore suggested that, although the source was overall in the soft 
% state, it may have temporarily returned to the HIMS. 
% Combined with our power color analysis, these results further suggest that 
% the state classification of this source may still require more detailed 
% investigation. 
% For MAXI J1820+070, the intensity of the initial HS observations first increases and then rapidly decreases, forming an additional structure in the HID. However, no obvious issues are found in its overall state classification. 
% For Swift J1727.8--1613, a flare state is also present, consistent with the results reported by \citet{yu2024timing}.}

\subsection{Hue vs Hardness}

\begin{figure*}
\centering
\includegraphics[width=0.7\textwidth]{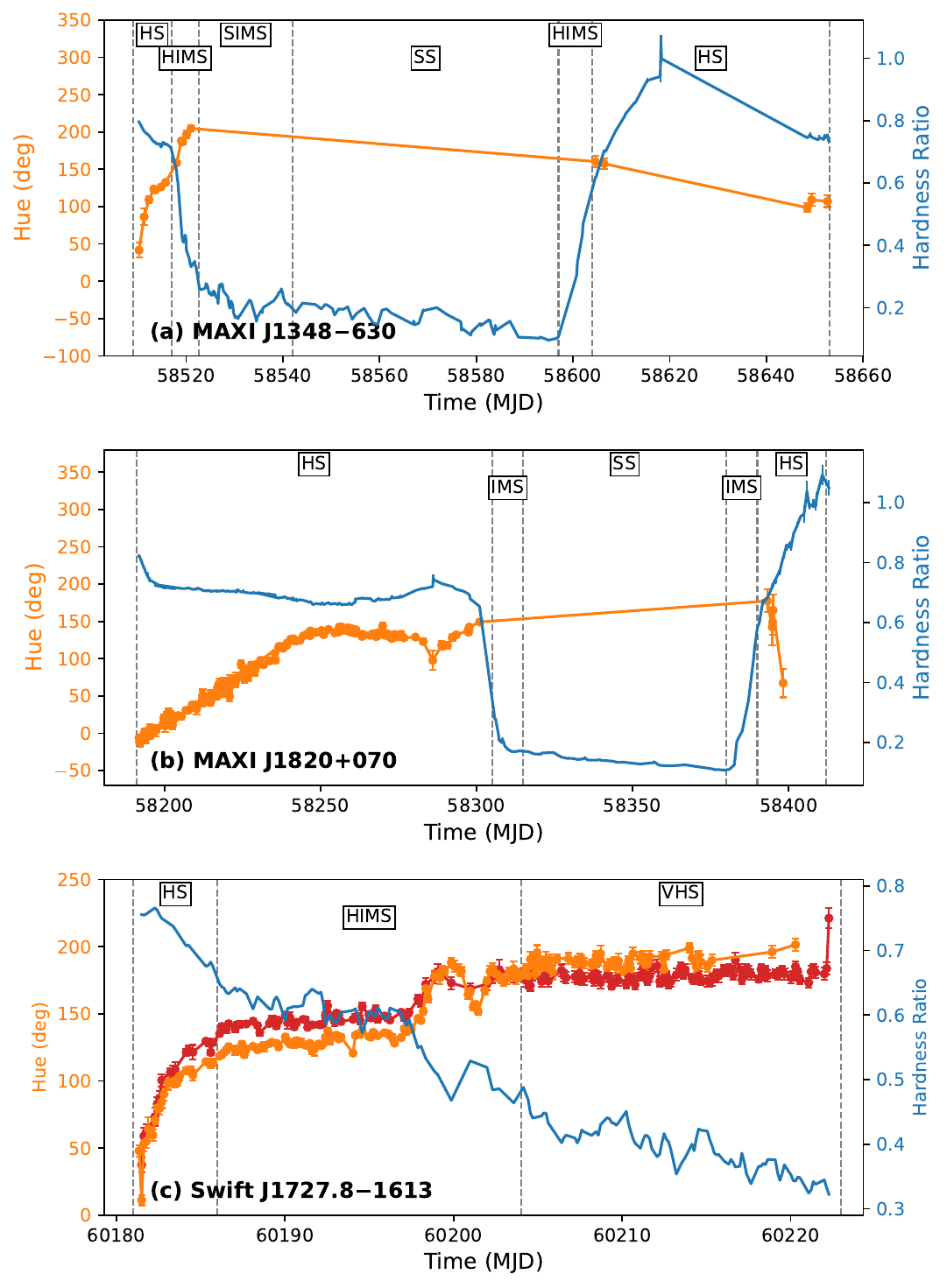}
\caption{
Evolutions of the hue (left; LE in red and ME in orange) and the spectral hardness (right; blue) for three black hole X-ray binaries.
Outburst states are adopted from \citet{carotenuto2026rapid}, \citet{peng2023possible}, \citet{cao2025spectral,liu2024broadbandxrayspectralproperties}.
}
% \caption{
% Evolution of the power spectral hue (blue points) and spectral hardness (red points) during the 2019 full outburst of MAXI~J1348$-$630.
% The outburst states are adopted from \citet{carotenuto2025full}.
% }
\label{fig:3hue}
\end{figure*}

\begin{figure}[t]
    \centering

    \begin{subfigure}[t]{0.45\textwidth}
        \centering
        \includegraphics[width=\linewidth]{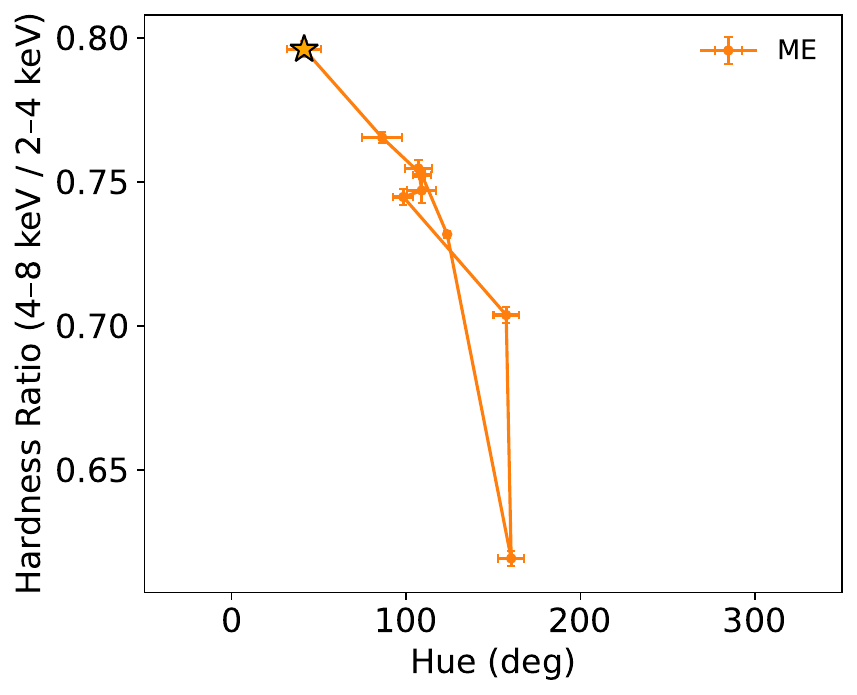}
        \caption{MAXI J1348--630/ME}
    \end{subfigure}
    \hfill
    \begin{subfigure}[t]{0.45\textwidth}
        \centering
        \includegraphics[width=\linewidth]{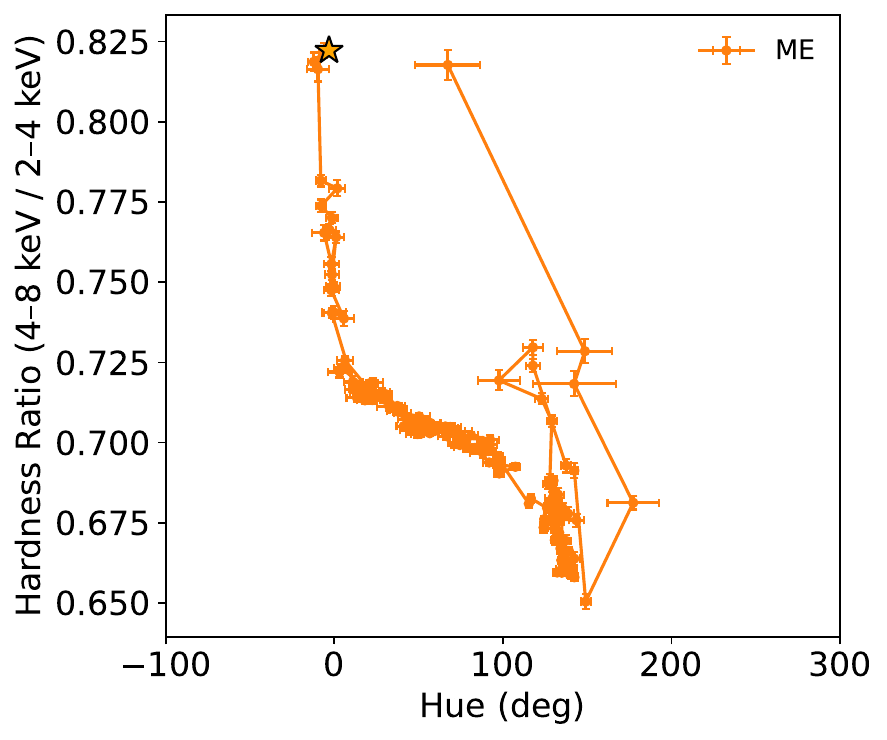}
        \caption{MAXI J1820+070/ME}
    \end{subfigure}

    % \vspace{-0.2cm}

    \begin{subfigure}[t]{0.45\textwidth}
        \centering
        \includegraphics[width=\linewidth]{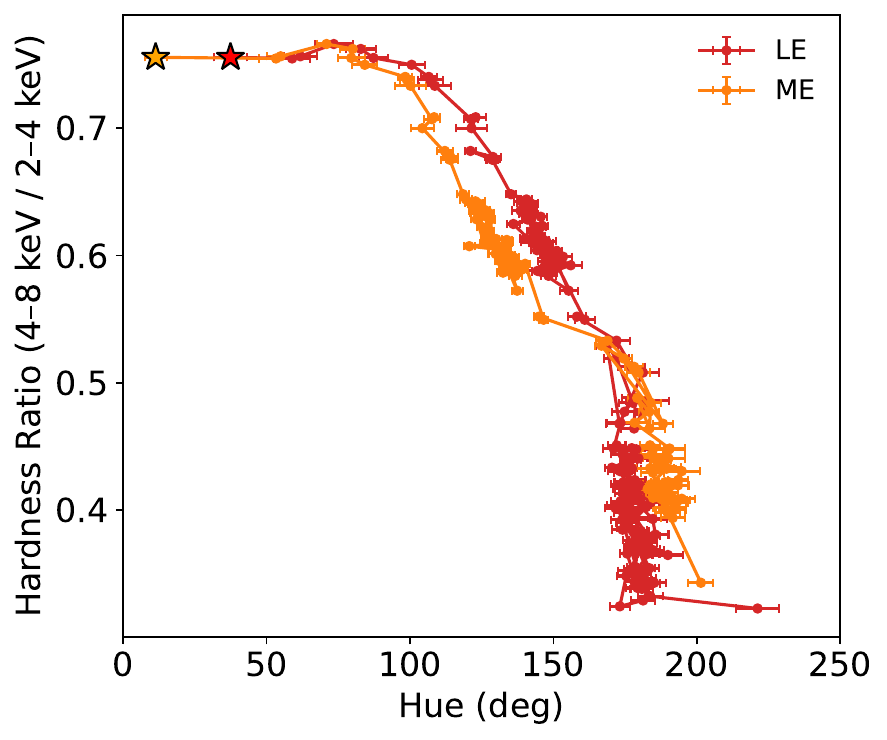}
        \caption{Swift J1727.8--1613/LE,ME}
    \end{subfigure}

    \caption{Relationships between the spectral hardness and the power spectral hue in the LE or ME band.
    The red and orange stars mark the starting points of their evolutionary tracks.
    }
    \label{fig:3huehr}
\end{figure}

%As mentioned in Section~\ref{sec:results}, we use {\it Insight}-HXMT observations to test the applicability of the {\it RXTE}-defined outburst state classification in the power colour--colour diagram.
% In this framework, each {\it RXTE}-defined hue region is associated with a specific outburst state, while the spectral hardness provides an independent diagnostic of the spectral properties.
The connection between the evolution of X-ray variability and spectral hardness during outbursts of black hole X-ray binaries has been extensively explored in previous studies.
In particular, \citet{lucchini2023variability} demonstrated that the variability could be a predictor for the hard-to-soft transition, given that the power spectral hue systematically evolves $\sim$10--40 days ahead of the canonical spectral state transition in GX 339-4.
%with a lead time of $\sim$10--40 days in full outbursts, where the source undergoes a complete transition from the HS to the SS and subsequently returns to the HS.
%In contrast, no such advance evolution of the hue was found in HS(hard state) outbursts, in which the source remains in the HS throughout the entire outburst \citep{lucchini2023variability}.
In Figure~\ref{fig:3hue} we present the hue and hardness evolutions during outbursts for our three sources.
Here the hardness was defined as the count rate ratio between 4--8 \,keV and 2--4 \,keV.
Because the evolutionary trajectories of different energy bands of MAXI J1348--630 and MAXI J1820+070 are similar in the power color-color diagrams (see Figure~\ref{fig:9pc}), we only show the ME results for clarity, whereas for Swift J1727.8--1613, we present both the LE and ME results. 
Interestingly, we found that for the VHS in Swift J1727.8--1613, although the power colors differ for different energy bands, their corresponding hue angles remain energy-independent.
In general, we found that, unlike in the case of GX 339--4, no time delays were detected between the hue and spectral evolutions.
Instead, they exhibit an approximate anti-correlation, except for the VHS of Swift J1727.8--1613, which has a stable hue regardless of variations in the hardness.
% But we note that this anti-correlation is not a simple linear.
% As shown in Figure~\ref{fig:3huehr}, they have different slopes at different epochs.
% Even for observations in the same HS with a comparable hardness, the hue evolution may manifest two distinct branches (see the HS in MAXI J1820+070 and Swift J1727.8--1613 in Figure~\ref{fig:3huehr}).
% In addition, when comparing preceding and lagging HS in MAXI J1820+070, their hue-hardness relations are not identical, which seems to be similar to the `q'-shaped hysteresis loop in the hardness-intensity diagrams \citep{fender2005unified}.
% This implies that relying solely on either the spectral or the timing signal is insufficient to determine outburst states, and combining information from both is required.
As shown in Figure~\ref{fig:3huehr}, the anti-correlation between hue and hardness is not a simple linear relation, but exhibits different slopes at different stages of the outburst evolution. 
In addition, for MAXI J1820+070, the tracks corresponding to the rising and decaying phases do not overlap in the hue--hardness diagram, showing clear differences between the two evolutionary branches. 
This behaviour is reminiscent of the  `q'-shaped hysteresis loop in the HID, where the source follows different paths during the rise and decay of the outburst. 
%\textcolor{blue}{(what do you mean "U-shaped"?)}
%\textcolor{red}{(Answer:the shape of figure 9a like a word 'U')}
These results suggest that relying solely on either spectral or timing information is insufficient to uniquely determine the outburst state, and that a combination of both is required.

% \begin{figure*}
% \centering
% \includegraphics[width=\textwidth]{figure/1820le_hue.pdf}
% \caption{
% Evolution of the power spectral hue (blue points) and spectral hardness (red points) during the 2018 full outburst of MAXI~J1820$+$070.
% The outburst states are adopted from \citet{peng2023possible}.
% }
% \label{fig:1820hue}
% \end{figure*}

% \begin{figure*}
% \centering
% \includegraphics[width=\textwidth]{figure/1727le_hue.pdf}
% \caption{
% Evolution of the power spectral hue (blue points) and spectral hardness (red points) during the 2023 full outburst of Swift~J1727.8$-$1613.
% The outburst states are adopted from \citet{cao2025spectral,liu2024broadbandxrayspectralproperties}.
% }
% \label{fig:1727hue}
% \end{figure*}

\section{CONCLUSIONS} \label{sec:summary}
Power color-color diagrams provide a phenomenological diagnostic to trace the evolution of the timing properties of black hole X-ray binaries during their outbursts.
This method was proposed by \cite{heil2015power} based on {\it RXTE} observations.
In this study, we tested its validity in {\it Insight}-HXMT data by comparing outburst states defined by timing and spectral studies with the positions in power color-color diagrams.
In addition, we expanded our investigation to multiple energy bands, encompassing the previously unaddressed higher-energy ranges of 10--30 \,keV and 30--80 \,keV.
Our main findings are summarized as follows:
\begin{itemize}

% \item Previous {\it RXTE}-defined hue regions, corresponding to different outburst states, should be modified when studying {\it Insight}-HXMT data (see Figure~\ref{fig:9pc}). 
% Nevertheless, for both the HS and the IMS, the {\it RXTE}-defined ranges are still roughly valid for {\it Insight}-HXMT.
% For the SS, due to large uncertainties, we can not well identify the position of MAXI J1820+070 in the power color-color diagram.
% For MAXI J1348--630, high frequency components are present in its soft state PDSs, which differs from other sources.
% As a result, its averaged point of power colors in the soft state is not consistent with the {\it RXTE}-defined region.

\item 
%For the HS and intermediate states, the power color points in the 
%three energy bands can generally be well associated with the corresponding 
%state regions defined by {\it RXTE}. 
%For MAXI J1348--630, the SS observations are located in the
%HIMS region of the power color-color diagram rather than
%in the SS region defined by {\it RXTE}. This displacement may be associated
%with the presence of type-A QPOs.
Overall, the hue regions established with \textit{RXTE} remain applicable to the multiband \textit{Insight}-HXMT observations, particularly in the hard and hard-intermediate states. In the soft and soft-intermediate states, however, the hue angles are generally difficult to constrain because of the weak variability. A few exceptions are found in MAXI J1348--630, where some hue angles deviate from the canonical {\it RXTE}-based locations. These deviations may be caused by averaging power spectra with variable shapes and, in the soft state, by the presence of a type-A QPO.

%these results suggest that the SS data points obtained with {\it Insight}-HXMT do not strictly coincide with the SS region defined by {\it RXTE}. 
%For MAXI J1348--630, this discrepancy can be attributed to the presence of a high-frequency hump in the SS power spectrum, which shifts the corresponding data points away from the {\it RXTE}-defined soft-state in the power colour--colour diagram. 
%For MAXI J1820+070, the {\it Insight}-HXMT results in the SS are associated with relatively large uncertainties, which limit the reliability of the comparison.

\item  In MAXI J1348--630 and MAXI J1820+070, the trajectories of different energy bands are basically similar. However, for Swift J1727.8--1613, the LE band exhibits a unique behavior.

\item  For the first time we reveal the VHS trajectory in the power color-color diagram, which in turn could be as a probe to verify the appearance of the VHS in the future.
This implies that one must be cautious when using hue angles to determine outburst states when only a limited number of observations are available.
For example, \citet{Ingram2024} classified the {\it IXPE} observations of Swift J1727.8--1613 as being in the HIMS, whereas the source was more likely in the VHS.

\item For Swift J1727.8--1613, in the VHS, the trajectory of LE is significantly different from those of ME and HE. We confirm that this is due to the presence of an additional low-frequency component in LE's power spectra.

\item We investigated the relationship between X-ray temporal variability, quantified by the power spectral hue, and spectral hardness during the outbursts of three black hole X-ray binaries observed by {\it Insight}-HXMT. 
For all sources studied in this work, we find that the evolution of hardness and hue tends to be anti-correlated, differing from the behavior reported by \citet{lucchini2023variability}.

\end{itemize}

\begin{acknowledgments}
We thank Bei You for helpful discussions. 
This work is supported by the National Natural Science Foundation of China under grant Nos. 12173103 and 12261141691. We acknowledge the use of data from the High Energy Astrophysics Science Archive Research Center (HEASARC), provided by NASA’s Goddard Space Flight Center, and from the {\it Insight}-HXMT mission, a project funded by the China National Space Administration (CNSA) and the Chinese Academy of Sciences (CAS).
\end{acknowledgments}

\bibliography{references}{}
\bibliographystyle{aasjournalv7}

\end{document}